\documentclass[aps,pre,onecolumn,showpacs,groupedaddress,eqsecnum]{revtex4}

\begin{document}

\preprint{1}

\title{Restoration of three-dimensional correlation function and structure factor 
from two-dimensional observations}


\author{Hiroo Totsuji}
\email[]{totsuji@elec.okayama-u.ac.jp}
\homepage[]{http://homepage3.nifty.com/totsuji/index2.html}
\affiliation{Graduate School of Natural Science and Technology
and 
Faculty of Engineering,
Okayama University,
Tsushimanaka 3-1-1, Okayama 700-8530, Japan}


\date{\today}

\begin{abstract}
The static pair correlation (distribution) function and the structure factor of particle distributions
in three-dimensional homogeneous isotropic systems are explicitly restored
from two-dimensional data observed in a thin slab sliced out from original systems.
While two-dimensional values for given thickness of the slab
are readily calculated from three-dimensional functions,
one would like to make the reverse in usual experiments
by, for example, scanning the slab perpendicularly.
Such a reconstruction is also possible even without scanning
and
three-dimensional pair correlation function and the structure factor are expressed
by two-dimensional data
in the form of an expansion with respect to the thickness of the slab.
As an application,
the behavior of the structure factor corresponding to the critical fluctuation is discussed.
These results are expected to be useful
when
three-dimensional systems are observed by the illumination of thin planar lasers. 
\end{abstract}


\maketitle


\section{Introduction}

Statistical properties of particle systems
are described by various correlation functions in the real space and corresponding spectra in the Fourier space.
Experimental observations of these functions and spectra are usually performed under limited conditions
and
it is necessary to evaluate them based on restricted information.
This paper is intended to contribute to one of such reconstructions,
the evaluation of the three-dimensional static pair correlation (distribution) function and the structure factor
from those obtained by two-dimensional observations.

We consider a three-dimensional system of particles
and
make a two-dimensional observation of a part of the system
which is cut out as a slice.
This situation is common for experiments where
particles are observed by the scattering of the thin planar laser beam or by the fluorescence induced by them.
For example,
in fine particle (dusty) plasma experiments,
the main target of observation is the behavior of fine particles (dusts)
and
their orbits are recorded by CCD cameras
through scattering of illuminating planar laser beams\cite{general}.
If the system is stationary and the timescale of the target phenomena is long enough,
one can obtain three-dimensional data
by scanning the beam perpendicularly to the sheet of the beam.
We may have, however, many cases where
the latter is not available. 

Our basic assumption is the statistical homogeneity and isotropy of the system.
We also assume
the threshold of detection by the light beam is adjusted so that 
all particles located within a given thickness are recorded
and
we have their two-dimensional positional data in the plane of the beam.
It is trivial that
values of the two-dimensional pair correlation function and the structure factor obtained by two-dimensional observations 
are determined by three-dimensional values.
We emphasize that,
under the assumption of the homogeneity and isotropy,
the reverse is also true:
{\it the three-dimensional pair correlation (distribution) function and the structure factor
can be reproduced from two-dimensional results.}
We have shown the inverse formula for the pair correlation (distribution) function
in the form of an expansion with respect to the thickness of the beam\cite{HT09a}.
We here derive the relations for the structure factors
(which have more mathematical aspects than those for correlation functions)
and
give a simple application of the result.

\section{correlation functions and structure factors in three and two dimensions}

We take the origin of coordinates at the center of the thickness of detection
with the $xy$-plane and the $z$-axis parallel and perpendicular to the beam sheet, respectively.
The system has $N$ particles in the volume $V=SL_z$
where
$S=L_xL_y$ is the area of the base parallel to the $xy$-plane 
and $L_z$ is the height in the $z$-direction.
The thickness and the volume of the sliced part detected by the beam are $b$ and $Sb$,
respectively.
We denote the number of particles in the observed domain by $N_{2d}$
which is related to $N$ by
\begin{equation}
N_{2d}=(b/L_z)N.
\end{equation}
We denote three- and two-dimensional coordinates
by ${\bf r}=(x, y, z)=({\bf R}, z)$ and ${\bf R}=(x, y)$, respectively. 
We also denote three- and two-dimensional wave numbers 
by ${\bf k}=(k_x, k_y, k_z)=({\bf K}, k_z)$ and ${\bf K}=(k_x, k_y)$, respectively.


The microscopic number densities in three and two dimensions,
$\rho({\bf r};\ 3d)$ and $\rho({\bf R};\ 2d)$, are defined by
\begin{equation}
\rho({\bf r};\ 3d) = \sum_{i=1}^N \delta({\bf r}-{\bf r}_i) = \sum_{i=1}^N \delta({\bf R}-{\bf R}_i)\delta(z-z_i)
\end{equation}
and
\begin{equation}
\rho({\bf R};\ 2d) = \int_{-b/2}^{b/2}dz \rho({\bf r};\ 3d),
\end{equation}
respectively.
The pair correlation functions in three and two dimensions,
$h(r;\ 3d)$ and $h(R;\ 2d)$,
 are defined respectively by
\begin{equation}
<\rho({\bf r};\ 3d) \rho({\bf r}';\ 3d)> 
= {N \over V}\delta({\bf r}-{\bf r}') + \left({N \over V}\right)^2[1+h(|{\bf r}-{\bf r}'|;\ 3d)]
\end{equation}
 and
\begin{equation}
<\rho({\bf R};\ 2d) \rho({\bf R}';\ 2d)> 
= {N_{2d} \over S}\delta({\bf R}-{\bf R}') + \left({N_{2d} \over S}\right)^2[1+h(|{\bf R}-{\bf R}'|;\ 2d)],
\end{equation}
where $<\ \ >$ denotes the statistical average.
These pair correlation functions are related to each other in the same way as the pair distribution functions\cite{HT09a}.
The two-dimensional correlation function is expressed by the three-dimensional one as
\begin{equation}\label{3dto2dpdf}
h(R;\ 2d) 
\approx h\left[ R\left(1+{1 \over 12}{b^2 \over R^2}-{1 \over 120}{b^4 \over R^4}\right);\ 3d \right]
+ {7 \over 1440}{b^4 \over R^4} R^2 {d^2h(R;\ 3d) \over dR^2},
\ \ \ \ \ {b \over R} < 1.
\end{equation}
Inversely,
the three-dimensional correlation function is expressed by the two-dimensional one as
\begin{equation}\label{2dto3dpdf}
h(r;\ 3d) 
\approx h\left[ r\left(1-{1 \over 12}{b^2 \over r^2}+{1 \over 720}{b^4 \over r^4}\right);\ 2d \right]
- {7 \over 1440}{b^4 \over r^4} r^2 {d^2h(r;\ 2d) \over dr^2},
\ \ \ \ \ {b \over r} < 1.
\end{equation}
We derive similar relations for the structure factors.

In three dimensions,
the Fourier component of the number density $\tilde{\rho}({\bf k};\ 3d)$ is defined by
\begin{equation}
\tilde{\rho}({\bf k};\ 3d) = \int_V d{\bf r} \exp(-i{\bf k}\cdot{\bf r}) \rho({\bf r};\ 3d)
\end{equation}
and
the inverse Fourier transform is given by
\begin{equation}
\rho({\bf r};\ 3d) = {1 \over V} \sum_{\bf k} \exp(i{\bf k}\cdot{\bf r}) \tilde{\rho}({\bf k};\ 3d),
\end{equation}
where
\begin{equation}
{\bf k} = 2 \pi \left({n_x \over L_x}, {n_y \over L_y}, {n_z \over L_z}\right),\ \ \ n_x, n_y, n_z=0, \pm1,\pm2,\dots.
\end{equation}
In two dimensions,
the Fourier component of the number density $\tilde{\rho}({\bf K};\ 2d)$ is defined by
\begin{equation}
\tilde{\rho}({\bf K};\ 2d) = \int_S d{\bf R} \exp(-i{\bf K}\cdot{\bf R}) \rho({\bf R};\ 2d)
\end{equation}
and
the inverse Fourier transform is given by
\begin{equation}
\rho({\bf R};\ 2d) = {1 \over S} \sum_{\bf K} \exp(i{\bf K}\cdot{\bf R}) \tilde{\rho}({\bf K};\ 2d),
\end{equation}
where
\begin{equation}
{\bf K} = 2 \pi \left({n_x \over L_x}, {n_y \over L_y}\right),\ \ \ n_x, n_y=0, \pm1,\pm2,\dots.
\end{equation}


In three and two dimensions,
the structure factors are defined respectively by
\begin{equation}
S({\bf k};\ 3d) = S(k;\ 3d) 
= {1 \over N} < |\tilde{\rho}({\bf k};\ 3d)|^2 >
\end{equation}
and
\begin{equation}
S({\bf K};\ 2d) = S(K;\ 2d) 
= {1 \over N_{2d}} < |\tilde{\rho}({\bf K};\ 2d)|^2 >.
\end{equation}
The correlation functions are related to structure factors by
\begin{equation}
h(r;\ 3d)={1 \over (2\pi)^3 (N/V)}\int d{\bf k} \exp(i{\bf k}\cdot {\bf r})[S(k;\ 3d)-1]
\end{equation}
and
\begin{equation}
h(R;\ 2d)= {1 \over (2\pi)^2 (N_{2d}/S)}\int d{\bf K} \exp(i{\bf K}\cdot {\bf R})[S(K;\ 2d)-1].
\end{equation}
Here,
the sums over wave numbers are rewritten into the integrals
assuming $V$ and $S$ are sufficiently large.
Similar limits will be taken in what follows
when possible without ambiguity.

\section{conversion of three-dimensional structure factor into two dimensions}


The density fluctuation in two dimensions $\tilde{\rho}({\bf K};\ 2d)$ is related to the three-dimensional one by
\begin{equation}
\tilde{\rho}({\bf K};\ 2d) 
= \int_S d{\bf R} \exp(-i{\bf K}\cdot{\bf R}) \int_{-b/2}^{b/2}dz \rho({\bf r};\ 3d)
= {b \over L_z} \sum_{k_z} \tilde{\rho}({\bf K}, k_z;\ 3d)
\left( {\sin (k_zb/2) \over k_zb/2}\right)
\end{equation}
and
$S(K;\ 2d)$ is calculated
from $S(k;\ 3d)$ as
\begin{eqnarray}
S(K;\ 2d) &=& {1 \over N_{2d}} < |\tilde{\rho}({\bf K};\ 2d)|^2 > \nonumber \\
&=& {b \over 2\pi} \int_{-\infty}^\infty dk_z S(k=(K^2+k_z^2)^{1/2};\ 3d) \left( {\sin (k_zb/2) \over k_zb/2}\right)^2.
\end{eqnarray}
We thus have
\begin{eqnarray}\label{3dto2d}
S(K;\ 2d)
&=& {b \over 2\pi} \int_{-\infty}^\infty dk_z S(k;\ 3d) \left( {\sin (k_zb/2) \over k_zb/2}\right)^2 \nonumber \\
&=& {b \over \pi} \int_K^\infty dk {k \over (k^2-K^2)^{1/2}}S(k;\ 3d) 
\left( {\sin [(k^2 -K^2)^{1/2}b/2] \over (k^2 -K^2)^{1/2}b/2}\right)^2.
\end{eqnarray}
Since
\[
{b \over 2\pi} \int_{-\infty}^\infty dk_z \left( {\sin (k_zb/2) \over k_zb/2}\right)^2
= 1,
\]
this relation is rewritten into the form
\begin{eqnarray}\label{3dto2dwith-1}
S(K;\ 2d) -1
&=& {b \over 2\pi} \int_{-\infty}^\infty dk_z [S(k;\ 3d)-1] \left( {\sin (k_zb/2) \over k_zb/2}\right)^2 \nonumber \\
&=& {b \over \pi} \int_K^\infty {dk k \over (k^2-K^2)^{1/2}} [S(k;\ 3d) - 1]
\left( {\sin [(k^2 -K^2)^{1/2}b/2] \over (k^2 -K^2)^{1/2}b/2}\right)^2.
\end{eqnarray}
When expanded with respect to $b$,
we have
\begin{equation}\label{3dto2db4}
S(K;\ 2d) - 1
= {b \over \pi} \int_K^\infty {dk k \over (k^2-K^2)^{1/2}} \left[
1 - {b^2 \over 12}  (k^2-K^2) + {b^4  \over 360} (k^2-K^2)^2
\right]
[S(k;\ 3d) - 1]
\end{equation}
to the order $b^4$.


The relation (\ref{3dto2d}) or (\ref{3dto2dwith-1}) should be derived 
also from the relation between correlation functions in the real space (\ref{3dto2dpdf}).
When we neglect terms proportional to $b^2$ or $b^4$ in (\ref{3dto2dpdf})
assuming $b$ is sufficiently small,
we have
\begin{equation}
h(R;\ 2d) \approx h(R;\ 3d).
\end{equation}
We then have
\begin{eqnarray}
S(K;\ 2d)-1 &=& 2 \pi {N_{2d} \over S} \int dR R J_0(KR)h(R;\ 2d) \nonumber \\
&\approx& 2 \pi {N_{2d} \over S} \int dR R J_0(KR)h(R;\ 3d) \nonumber \\
&=& {b \over \pi} \int_0^\infty dR R J_0(KR) \int_0^\infty dk k^2 {\sin(kR) \over kR}[S(k;\ 3d)-1].
\end{eqnarray}
Noting that
\begin{equation}
\int_0^\infty dR R {\sin (kR) \over kR} J_0(KR) = { 1 \over k(k^2-K^2)^{1/2}} \theta (k-K),
\end{equation}
we have
\begin{equation}\label{3dto2dwith-1b0}
S(K; 2d)-1 = {b \over \pi}\int_K^\infty {dk k \over (k^2-K^2)^{1/2}}[S(k;\ 3d)-1],
\end{equation}
which is the $b \rightarrow 0$ limit of (\ref{3dto2dwith-1})
(Since $S(k;\ 3d) \rightarrow 1$ as $k \rightarrow \infty$,
we have to treat this limiting value of $S(k;\ 3d)$ separately
for the convergence of the integral).

\section{Inverse relations for structure factors}

In the real space,
we have the inverse relation (\ref{2dto3dpdf}) for the correlation function.
In the small $b$ limit,
we have
\begin{equation}
h(r;\ 3d) \approx h(r;\ 2d).
\end{equation}
This gives
\begin{eqnarray}
S(k;\ 3d)-1 
&=& {N \over V} 4\pi \int_0^\infty dr r^2 {\sin (kr) \over kr} h(r;\ 3d) \nonumber \\
&\approx& {N \over V} 4\pi \int_0^\infty dr r^2 {\sin (kr) \over kr} h(r;\ 2d) \nonumber \\
&=& {2 \over b} \int_0^\infty dr r^2 {\sin (kr) \over kr} \int_0^\infty dK KJ_0(Kr)[S(K;\ 2d)-1] \nonumber \\
&=& - {2 \over b} \int_0^\infty dr r {\sin (kr) \over kr} \int_0^\infty dK KJ_1(Kr) {d \over dK} [S(K;\ 2d)-1].
\end{eqnarray}
Noting that
\begin{equation}
\int_0^\infty dr r {\sin (kr) \over kr} J_1(Kr) = {1 \over K(K^2 - k^2)^{1/2}} \theta(K-k),
\end{equation}
we have
\begin{equation}\label{2dto3dwith-1b0}
S(k;\ 3d)-1= - {2 \over  b} \int_k^\infty dK {1 \over (K^2 - k^2)^{1/2}} {d \over dK} [S(K;\ 2d)-1].
\end{equation}
Equation (\ref{2dto3dwith-1b0}) is the inverse of the relation (\ref{3dto2dwith-1b0})
as is also directly checked by substituting (\ref{2dto3dwith-1b0}) into (\ref{3dto2dwith-1b0}):
\begin{eqnarray}\label{inverse}
& &{b \over \pi}\int_K^\infty {dk k \over (k^2-K^2)^{1/2}}
\left[- {2 \over b} \int_k^\infty dK' {1 \over (K'^2 - k^2)^{1/2}} {d \over dK'} [S(K';\ 2d)-1]\right] \nonumber \\
&=& - {2 \over \pi} \int_K^\infty dK' {d \over dK'} [S(K';\ 2d)-1] \int_K^{K'} {dk k \over (k^2-K^2)^{1/2}}
{1 \over (K'^2 - k^2)^{1/2}} \nonumber \\
&=& -  \int_K^\infty dK' {d \over dK'} [S(K';\ 2d)-1] = S(K;\ 2d)-1.
\end{eqnarray}

Equations (\ref{3dto2dwith-1b0}) and (\ref{2dto3dwith-1b0}) give the transformations 
from $S(k;\ 3d)$ to $S(K;\ 2d)$ and vise versa
in the lowest order in $b$.
While (\ref{3dto2d}) or (\ref{3dto2dwith-1}) gives exact values of $S(K;\ 2d)$
including higher order terms,
the purpose of experiments is to obtain values of $S(k;\ 3d)$ from those of $S(K;\ 2d)$.
Let us now derive higher order terms in the inverse formula for $S(k;\ 3d)$.

Since the inverse of (\ref{3dto2dwith-1}) is given by (\ref{2dto3dwith-1b0})
in the lowest order,
we put
\begin{equation}\label{formalsolution}
S(k;\ 3d) - 1
= {1 \over b}\left[
- 2 \int_k^\infty {dK \over (K^2-k^2)^{1/2}} {d \over dK}[S(K;\ 2d) - 1]
+ b^2\Delta^{(2)}(k) + b^4\Delta^{(4)}(k) \right]
\end{equation}
and determine $\Delta^{(2)}(k)$ and $\Delta^{(4)}(k)$
so as to satisfy (\ref{3dto2db4}).
Substituting (\ref{formalsolution}) into (\ref{3dto2db4}) 
and noting (\ref{inverse}),
we have
\begin{eqnarray}\label{invb2}
& &\int_K^\infty  {dk k \over (k^2-K^2)^{1/2}} \Delta^{(2)}(k) \nonumber \\
= &-& {1 \over 6} \int_K^\infty dk k (k^2-K^2)^{1/2} 
\int_k^\infty {dK'  \over (K'^2-k^2)^{1/2}}{d \over dK'}[S(K';\ 2d) - 1]
\end{eqnarray}
and
\begin{eqnarray}\label{invb4}
& & \int_K^\infty {dk k \over (k^2-K^2)^{1/2}} \Delta^{(4)}(k)
=
{1 \over 12}\int_K^\infty dk k (k^2-K^2)^{1/2} \Delta^{(2)}(k) \nonumber \\
&+&
{1 \over 180}\int_K^\infty dk k (k^2-K^2)^{3/2}
\int_k^\infty {dK' \over (K'^2-k^2)^{1/2}} {d \over dK'}[S(K';\ 2d) - 1] 
\end{eqnarray}
in the orders $b^2$ and $b^4$, respectively.

Rewriting the right-hand side of (\ref{invb2}) as
\begin{eqnarray}
&-& {1 \over 6} \int_K^\infty dK' {d \over dK'}[S(K';\ 2d) - 1] 
\int_K^{K'} dk k {(k^2-K^2)^{1/2} \over (K'^2-k^2)^{1/2}} \nonumber \\
= &-& {\pi \over 24} \int_K^\infty dK' (K'^2-K^2){d \over dK'}[S(K';\ 2d) - 1]
= {\pi \over 12} \int_K^\infty dK' K' [S(K';\ 2d) - 1]
\end{eqnarray}
and
noting that 
the inverse of (\ref{3dto2dwith-1b0}) is given by (\ref{2dto3dwith-1b0}),
we can solve (\ref{invb2}) for $\Delta^{(2)}(k)$ 
in the form
\begin{eqnarray}\label{Delta^2}
\Delta^{(2)}(k) &=& - {2 \over  b} \int_k^\infty {dK \over (K^2 - k^2)^{1/2}} {d \over dK} 
\left[
{b \over 12} \int_K^\infty dK' K' [S(K';\ 2d) - 1] \right] \nonumber \\
&=&  {1 \over 6} \int_k^\infty {dK \over (K^2 - k^2)^{1/2}} K [S(K;\ 2d) - 1].
\end{eqnarray}
Substituting (\ref{Delta^2}) into (\ref{invb4}),
we have
\[
\int_K^\infty {dk k \over (k^2-K^2)^{1/2}} \Delta^{(4)}(k)
\]
\[
=
{1 \over 72} \int_K^\infty dk k (k^2-K^2)^{1/2} 
\int_k^\infty {dK' \over (K'^2 - k^2)^{1/2}} K' [S(K';\ 2d) - 1]
\]
\[
+ {1 \over 180} \int_K^\infty dk k (k^2-K^2)^{3/2}
\int_k^\infty {dK' \over (K'^2-k^2)^{1/2}} {d \over dK'}[S(K';\ 2d) - 1].
\]
The right-hand side is rewritten as
\[
{1 \over 72} \int_K^\infty dK' K' [S(K';\ 2d) - 1] 
\int_K^{K'} dk k {(k^2-K^2)^{1/2} \over (K'^2 - k^2)^{1/2}} 
\]
\[
+ {1 \over 180} \int_K^\infty dK' {d \over dK'}[S(K';\ 2d) - 1]
\int_K^{K'} dk k  {(k^2-K^2)^{3/2} \over (K'^2-k^2)^{1/2}} 
\]
\[
=
- {\pi \over 1440} \int_K^\infty dK' (K'^2-K^2) K' [S(K';\ 2d) - 1]
\]
and
(\ref{invb4}) is solved for $\Delta^{(4)}(k)$ in the form
\begin{eqnarray}\label{Delta^4}
\Delta^{(4)}(k) &=& {1 \over 720} \int_k^\infty {dK \over (K^2 - k^2)^{1/2}} {d \over dK} 
\int_K^\infty dK' (K'^2-K^2) K' [S(K';\ 2d) - 1] \nonumber \\
&=& - {1 \over 360} \int_K^\infty dK' K' [S(K';\ 2d) - 1] \int_k^{K'}  {dK K \over (K^2 - k^2)^{1/2}} \nonumber \\
&=& - {1 \over 360} \int_k^\infty dK K (K^2 - k^2)^{1/2} [S(K;\ 2d) - 1].
\end{eqnarray}
Substituting (\ref{Delta^2}) and (\ref{Delta^4}) into (\ref{formalsolution}),
we have finally
\begin{equation}\label{2d3dexpansion}
S(k;\ 3d) - 1
= 
\int_k^\infty {dK \over (K^2-k^2)^{1/2}} 
\left[-{2 \over b} {d \over dK} + {b \over 6} K - {b^3 \over 360} K (K^2-k^2) \right][S(K;\ 2d) - 1]
\end{equation}
as the inverse of (\ref{3dto2db4}).

\section{Application}

One of remarkable phenomena
observed in the static structure factors
may be the density fluctuations near the critical point:
When the critical point is approached,
an enhancement and eventual divergence are expected in the spectrum of long wavelength density fluctuations\cite{LLSP}.
A possibility of the critical point has been pointed out in fine particle plasmas\cite{HT06,HT08a09b}
where
particle orbits are usually observed by thin laser beams
and
the behavior of density fluctuations near the critical point has been calculated\cite{HT08a09b}.
We here show
how such a divergence looks in two-dimensional images.

Let us assume that
the three-dimensional structure factor shows the critical behavior at long wavelengths in the form
\begin{equation}
S(k;\ 3d) \sim {1 \over c^2 + (k/k_0)^2},\ \ \ \ k \rightarrow 0
\end{equation}
with $c^2 \rightarrow 0$ at the critical point.
When observed by the laser beam of thickness $b$,
the two-dimensional structure factor takes the form given by (\ref{3dto2d}).
If $b$ is sufficiently small,
we have
\begin{equation}
S(K;\ 2d)
\sim 
{b k_0 \over 2}{1 \over (c^2 + K^2/k_0^2)^{1/2}}.
\end{equation}
The two-dimensional structure factor observed by the laser beam of thickness $b$ has a {\it weaker divergence}
in proportion to $c$ as
\begin{equation}
S(K;\ 2d) \sim {b k_0 \over 2c},\ \ \ \ K \rightarrow 0
\end{equation}
in comparison with the (true) three-dimensional one
\begin{equation}
S(k;\ 3d) \sim {1 \over c^2},\ \ \ \ k \rightarrow 0,
\end{equation}
while
the wave number characterizing the behavior in $K$-space is $k_0$,
the same as the three-dimensional behavior.

At long wavelengths,
two-dimensional results takes the form
\begin{equation}\label{2dfitting}
{1 \over S(K;\ 2d) ^2} \sim C^2_1 + K^2/K_0^2
\end{equation}
with
\begin{equation}
C^2_1 = \left({2 \over b k_0}\right)^2 c^2
\end{equation}
and
\begin{equation}
K_0 = { b k^2_0  \over 2}.
\end{equation}
We thus have
\begin{equation}
c^2 = {bK_0 \over 2} C^2_1
\end{equation}
and
\begin{equation}
k_0^2 = 2 {K_0 \over b}.
\end{equation}
These equations give $c^2$ and $k_0$ of the true structure factor $S(k;\ 3d)$
in terms of the values obtained by two-dimensional observations,
$C^2_1$ and $K_0$,
and the (known) value of the beam thickness $b$.

\section{conclusion}

In addition to those for the correlation (distribution) functions,
the relations between the three- and two-dimensional structure factors
are given.
It is obvious that,
when the three-dimensional structure factor is known for homogeneous and isotropic systems,
we can calculate the values of two-dimensional structure factor observed in a slab sliced out from three-dimensional systems.
It is shown that
the reverse is also possible:
The three-dimensional structure factor is expressed by the values of two-dimensional one
in the form of an expansion with respect to the thickness of the slab.
The results is applied to the case of critical density fluctuations expected in the structure factor near the critical point.
Characteristic parameters of critical fluctuations are related to those observed in two-dimensional structure factor,
enabling estimation of the former by two-dimensional observations.

\begin{acknowledgments}
The author thanks Drs. H. M. Thomas, K. Takahashi, and S. Adachi for information on fine particle experiments.
This work was supported by the Grant-in-Aid for Scientific Research (C) 21540512
from Japan Society for the Promotion of Science.
\end{acknowledgments}



\end{document}